\documentclass[aps,prl,twocolumn,superscriptaddress,showpacs]{revtex4}

\usepackage{amssymb}
\usepackage{amsmath}
\usepackage{graphicx}
\usepackage{natbib}
\usepackage{nicefrac}
\usepackage{multirow}
\usepackage{epstopdf}
\usepackage{float}
\usepackage{color}

\begin{document}

\title{Investigation of Mobility Limiting Mechanisms in Undoped Si/SiGe Heterostructures}
\author{X. Mi}
\affiliation{Department of Physics, Princeton University, Princeton, NJ 08544, USA}
\author{T. M. Hazard}
\affiliation{Department of Physics, Princeton University, Princeton, NJ 08544, USA}
\author{C. Payette}
\affiliation{Department of Physics, Princeton University, Princeton, NJ 08544, USA}
\author{K. Wang}
\altaffiliation{Present Address: Department of Physics, Harvard University, Cambridge, MA 02138, USA}
\affiliation{Department of Physics, Princeton University, Princeton, NJ 08544, USA}
\author{D. M. Zajac}
\affiliation{Department of Physics, Princeton University, Princeton, NJ 08544, USA}
\author{J. V. Cady}
\affiliation{Department of Physics, Princeton University, Princeton, NJ 08544, USA}
\author{J. R. Petta}
\affiliation{Department of Physics, Princeton University, Princeton, NJ 08544, USA}
\affiliation{Department of Physics, University of California, Santa Barbara, CA 93106, USA}

\pacs{73.21.Fg, 73.21.La, 85.30.De}

\begin{abstract}
We perform detailed magnetotransport studies on two-dimensional electron gases (2DEGs) formed in undoped Si/SiGe heterostructures in order to identify the electron mobility limiting mechanisms in this increasingly important materials system. By analyzing data from 26 wafers with different heterostructure growth profiles we observe a strong correlation between the background oxygen concentration in the Si quantum well and the maximum mobility. The highest quality wafer supports a 2DEG with mobility $\mu$ = 160,000 cm$^{2}$/Vs at a density $n$ $=$ 2.17 $\times 10^{11}/$cm$^{2}$ and exhibits a metal-to-insulator transition at a critical density $n_{c}$ = $0.46 \times 10^{11}/$cm$^2$. We extract a valley splitting $\Delta_\text{v}$ $\sim150$ $\mu$eV at a magnetic field $B$ = 1.8 T. These results provide evidence that undoped Si/SiGe heterostructures are suitable for the fabrication of few-electron quantum dots.
\end{abstract}

\maketitle
\section{Introduction}
The development of silicon quantum devices has gained considerable momentum due to reports of quantum coherence times ($T_2$) as long as 39 minutes \cite{Thewalt_39mins}. Its naturally abundant isotope, $^{28}$Si, carries zero nuclear spin, reducing hyperfine-induced dephasing due to fluctuations of the nuclear spin bath \cite{Erik_QIP,Petta_Science,Petta_RevMod}. Its small spin-orbit coupling is also beneficial for spin qubits \cite{Si_Spin_Transport,PhysRevB.77.073310}. Following work in GaAs quantum dots, early experimental efforts were made towards fabricating Si quantum dots in modulation-doped Si/SiGe heterostructures, where the n-type dopant layer is separated from the Si quantum well (QW) by a setback distance ranging from 5 to 20 nm \cite{UW_Early_Doped_SiGe,Eriksson_NatPhys_2008,SiGe_Doped_Topgated,Payette_APL}. Doped devices encountered challenges when operating in the few-electron regime, displayed hysteresis in gate voltage sweeps \cite{Payette_APL}, and sometimes suffered from leakage between the 2DEG and depletion gates \cite{UW_Early_Doped_SiGe,SiGe_Doped_Topgated}.

It is now widely accepted that the elimination of the n-type dopant layer decreases the Coulomb disorder in the QW, and reduces hysteresis and gate leakage \cite{Sandia_SiGe_Hall,HRL_SiGeAPL}. Recent experiments focusing on quantum dots made in undoped Si/SiGe QWs \cite{HRL_SiGeAPL,Sandia_SiGe_Hall,Ke_PRL,HRL_Nature2012} have consistently reached the single-electron regime and demonstrated inhomogeneous spin dephasing times $T_2^*$ = 360 ns in naturally abundant Si, a substantial increase compared to GaAs spin qubits \cite{Petta_Science,HRL_Nature2012}. Further improvement of the Si/SiGe QW system may be feasible if the remaining mobility limiting mechanisms are clearly identified \cite{Monroe_Mobility,DasSharma_SiGe_PRB2005,Gold_JAP_SiGe}.

The dominant scattering sources can be identified from measurements of the carrier mobility $\mu$ as a function of 2DEG charge density $n$, as well as measurements of the quantum lifetime $\tau_q$ \cite{Monroe_Mobility,DasSharma_PRB1985}. For example, scattering from remote impurities \cite{Monroe_Mobility} is predicted to result in a power-law dependence $\mu \propto n^{1.5}$. Such experiments have been extensively performed for GaAs/AlGaAs heterostructures \cite{Shayegan_Early_GaAs,Jiang_GaA_Mobility,Bockelmann_GaA,Coleridge_InterSubband,Coleridge_SmallAngle,GaA_APL1997}, GaN/AlGaN heterostructures \cite{APL_GaN_2004,IOP_GaN_2007}, Si MOSFETs \cite{Si_MOSFET_Scattering} and doped Si/SiGe heterostructures \cite{DopedSiGe_APL_1995,Sugii_DopedSiGe_1998,Sturm_APL2012}. The conclusions reached in doped Si/SiGe heterostructures are not directly applicable to undoped structures, and similar measurements on undoped Si/SiGe heterostructures are scarce. One recent work reports a record-high mobility of $2 \times 10^6$ cm$^2$/Vs at a temperature $T$ = 0.3 K and $n = 1.4 \times 10^{11}/$cm$^{2}$  in an undoped Si/SiGe QW and identifies background impurity charges and interface roughness at the QW as the main mobility-limiting mechanisms \cite{Tsui_APL2012_HighSiGeMobility}. However, the 526 nm thick spacer layer used in this work is much too large to allow the tight electrostatic confinement that is needed for few-electron quantum dot devices \cite{Erik_QIP}. Li \textit{et al.} investigated a heterostructure with a spacer layer thickness of 60 nm and deduced that remote impurity charges at the Al$_2$O$_3$/Si interface limit the mobility \cite{Sturm_APL2012}.

To thoroughly investigate the mobility limiting mechanisms in undoped Si/SiGe QWs, we report a series of systematic magnetotransport measurements in the density range 0.5 -- 8.0 $\times$  $10^{11}/$cm$^{2}$ at temperatures from 0.35 -- 4.2 K. By examining 26 different heterostructure growth profiles, we identify a strong correlation between background oxygen concentration in the QW and maximum mobility. These results indicate that significant enhancements in Si/SiGe mobility might be obtained through more careful control of background contamination during heterostructure growth.

\begin{figure}
\centering
\includegraphics[width=\columnwidth]{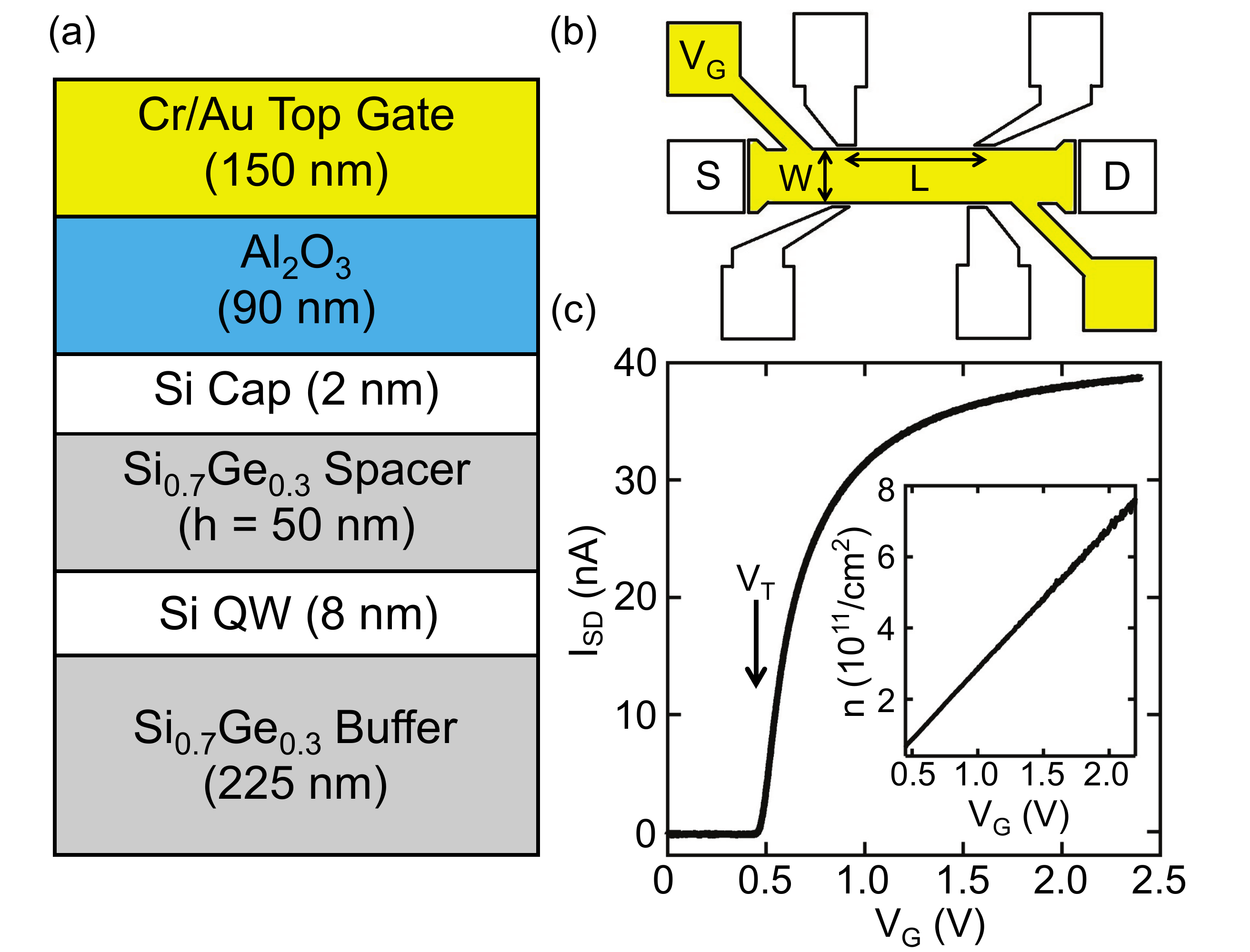}
\caption{(a) Heterostructure growth profile. (b) Top view of the Hall bar device. The gold region marks the area covered by the top gate. The other six square pads are $^{15}$P implanted regions which form Ohmic contacts to the electron gas. The Hall bar dimensions are $W$ = 170 $\mu$m and $L$ = 375 $\mu$m. (c) A typical ``turn-on'' curve of the device at $T = $ 0.35 K showing the source-drain current $I_\text{SD}$ as a function of $V_{\text G}$. The threshold voltage for non-zero $I_\text{SD}$ is $V_{\text T} = 0.45$ V. Inset shows $n$ for $V_{\text G} > V_{\text T}$, before saturation. Dimensions and data shown in (a) and (c) are for Wafer No.\ 16.}
\label{fig:1}
\end{figure}

\section{Silicon Germanium Heterostructures}
The samples examined here were grown at Lawrence Semiconductor Research Laboratory using chemical vapor deposition. 26 Si/SiGe heterostructures are investigated in order to distinguish mobility limiting mechanisms that are related to the growth profile [Fig.~\ref{fig:1}(a)] from those that are related to background impurities. Relaxed buffers of Si$_{1-x}$Ge$_x$ are first grown on Si substrates, varying $x$ from 0 to 0.3 over a thickness of 3 $\mu$m. A 1 $\mu$m thick layer of Si$_{0.7}$Ge$_{0.3}$ is grown on the virtual substrate before it is polished. The wafers are completed by growing a 225 nm thick Si$_{0.7}$Ge$_{0.3}$ layer, followed by a Si QW which is strained through the Si/Si$_{0.7}$Ge$_{0.3}$ lattice mismatch, a Si$_{0.7}$Ge$_{0.3}$ spacer layer and a protective Si cap. We investigate heterostructures with Si cap thicknesses of 2 nm and 4 nm, Si$_{0.7}$Ge$_{0.3}$ spacer layer thicknesses $h$ = 20 nm, 30 nm, 40 nm and 50 nm, and Si QW thicknesses of 5 nm, 8 nm and 11 nm.

Hall bars are fabricated on each of the 26 wafers, with the geometry shown in Fig.~\ref{fig:1}(b). We first use atomic layer deposition to grow an Al$_2$O$_3$ gate dielectric on top of the Si cap. We then evaporate Cr/Au on top of the Al$_2$O$_3$ to form a top gate. A positive dc bias is applied to the top gate to accumulate electrons in the QW and a 0.1 mV, 17 Hz ac voltage excitation is applied between the ohmic contacts marked with $S$ and $D$ in Fig.~\ref{fig:1}(b). The longitudinal voltage, $V_\text{xx}$, and the Hall voltage, $V_\text{xy}$, are simultaneously measured as a function of magnetic field $B$ using standard ac lock-in techniques. The source-drain current, $I_\text{SD}$, is also measured using a current-to-voltage pre-amplifier and a lock-in amplifier. Employing a fixed voltage source in place of a fixed current source prevents a destructively large voltage from being applied across the sample at low electron densities where the longitudinal resistance is large. The 2D longitudinal resistivity, $\rho_\text{xx} = (V_\text{xx} / I_\text{SD}) (W/L)$, and Hall resistivity, $\rho_\text{xy} = (V_\text{xy} / I_\text{SD})$ are calculated from the measured voltages and currents. We have confirmed that the voltage bias is not heating the sample by comparing measurements of $\rho_\text{xx}$ and $\rho_\text{xy}$ at different excitation voltages. Density, $n$, and mobility, $\mu$, of carrier electrons are calculated according to the Hall formulas $n = B /( e \rho_\text{xy})$ and $\mu = (1 / B) (\rho_\text{xy} / \rho_\text{xx})$.

Figure \ref{fig:1}(c) displays a typical ``turn-on'' curve of the Hall bar devices. Zero current flow is observed below a threshold top gate voltage $V_\text{T} = 0.45$ V. For $V_\text{G} > V_\text{T}$, current starts to flow and we observe a linear increase in $n$ with a slope of $dn / dV_\text{G} = 3.96 \times 10^{11}/$cm$^{2}/$V. At $V_\text{G} > 2.5$ V, the electron density is fixed at a constant value of 8.0 $\times$ $10^{11}/$cm$^{2}$. Such behavior can be understood as follows: We define $\Delta E_\text{c} = e(V_\text{G} - V_\text{T})$ to be the difference between the chemical potential of the 2DEG electrons and the conduction band minimum of 2DEG electrons, where $e = 1.60 \times 10^{-19}$ C is the magnitude of the electron charge. For $V_\text{G} < V_\text{T}$, $\Delta E_\text{c} < 0$ and the chemical potential of the 2DEG electrons lies below the conduction band minimum. As a result, the QW is depleted of carriers. When $V_\text{G} > V_\text{T}$, $\Delta E_\text{c} > 0$ and the conduction band becomes populated. The Hall bar behaves like a parallel plate capacitor at these voltages. Using relative permittivities of $\epsilon_r = 9$ for Al$_2$O$_3$ and $\epsilon_r = 13.1$ for Si$_\text{0.7}$Ge$_\text{0.3}$, we calculate $dn / dV_\text{G} = 4.00 \times 10^{11}/$cm$^{2}/$V, which is within 1 \% of the experimental value. At even higher values of $V_\text{G}$ (data not shown), electrons start to accumulate at the Al$_2$O$_3$/Si cap interface, screening the QW from any further increase in $V_\text{G}$. This causes a saturation of the electron density at a constant value of 8.0 $\times$ $10^{11}/$cm$^{2}$ for $V_\text{G} > 2.5$ V. Lu \textit{et al.} have investigated such saturation behavior in detail \cite{Tsui_APL2011_MaxDensity}.

\section{Characterization at $T$ = 4.2 K}

\begin{figure}
\centering
\includegraphics[width=\columnwidth]{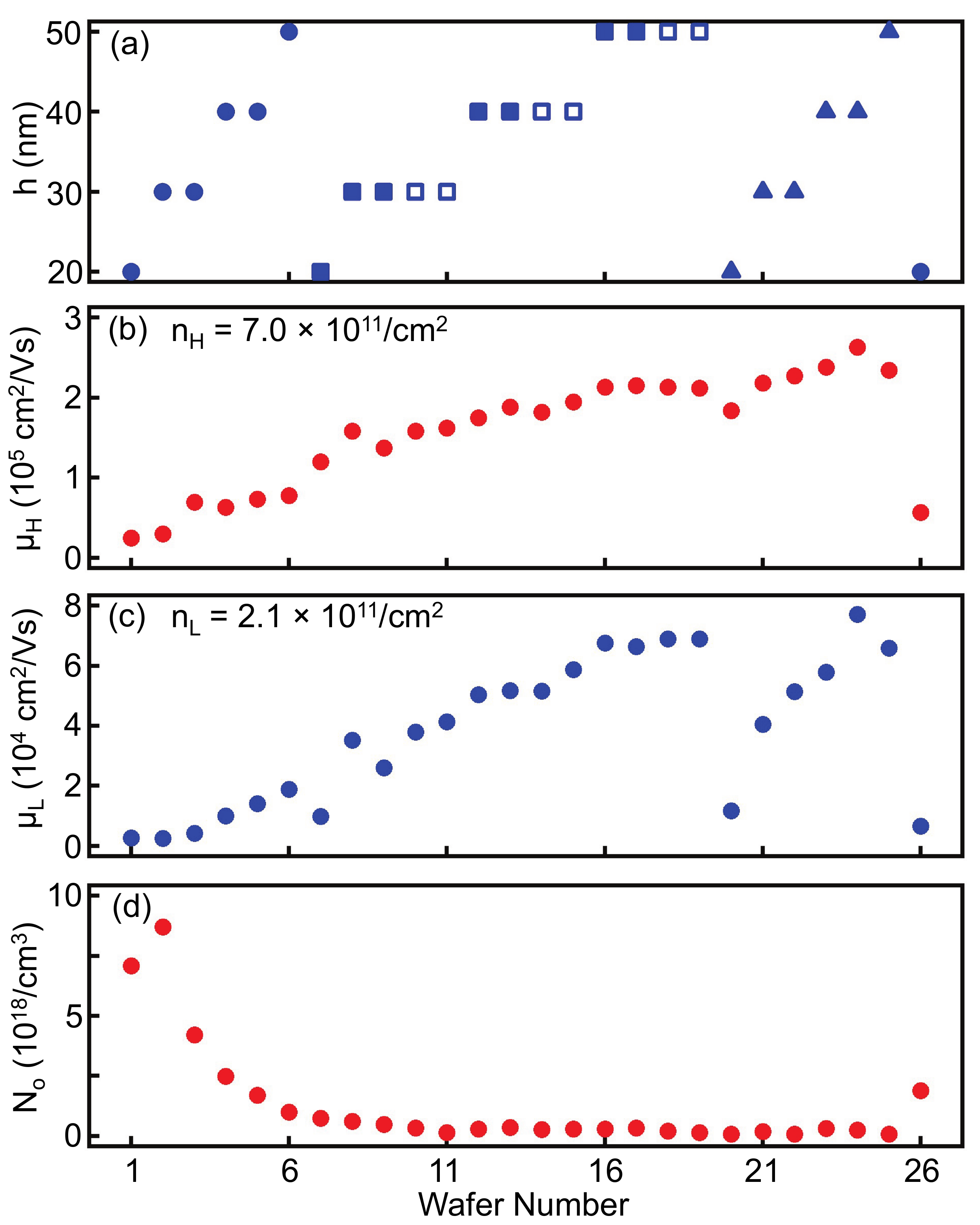}
\caption{(a) SiGe spacer layer thickness, $h$, for wafers 1--26. Wafers with solid (hollow) symbols have a 2 nm (4 nm) thick Si cap. QW thicknesses are represented by symbol shapes. Circles: 5 nm, rectangles: 8 nm, triangles: 11 nm. (b) $\mu_{\text H}$ is the $T$ = 4.2 K mobility at $n_{\text H}$ = 7 $\times$ 10$^{11}/$cm$^{2}$. (c) $\mu_{\text L}$ is the $T$ = 4.2 K mobility at $n_{\text L}$ = 2.1 $\times$ 10$^{11}/$cm$^{2}$. (d) $N_\text{o}$ is the concentration of oxygen atoms inside the QW, obtained from SIMS data. SIMS is limited to measuring oxygen concentrations above 1 $\times$ $10^{17}/$cm$^{3}$.}
\label{fig:2}
\end{figure}

Hall bars are first measured at $T$ = 4.2 K and $B =$ 0.1 T, below the onset of Shubnikov-de Haas (SdH) oscillations. Figure 2(a) shows the spacer layer thickness $h$ for each of the 26 wafers, along with the Si cap thickness and QW width. Recent studies of undoped Si/SiGe structures have shown that remote impurity scattering typically dominates in the low electron density regime, whereas both remote impurities and interface roughness dominate at higher electron densities \cite{Sturm_APL2012,Tsui_APL2012_HighSiGeMobility}. It is therefore helpful to examine electron mobilities at both density regimes. Corresponding electron mobilities are plotted in Fig.\ 2(b) for a high electron density $n_{\text H}$ = 7.0 $\times$ $10^{11}/$cm$^{2}$ and in Fig.\ 2(c) for a low electron density $n_{\text L}$ = 2.1 $\times$ $10^{11}/$cm$^{2}$. Surprisingly, both the low and high density mobilities show a nearly monotonic increase with wafer number, despite the large variation in heterostructure parameters throughout this series of wafers. On top of this trend, abrupt dips in the mobility are observed at Wafer No.\ 20 and 26.

Secondary ion mass spectrometry (SIMS) analysis was performed on each wafer to better understand the increase in mobility as a function of wafer number. These data sets are included in the supplemental material \cite{Supplement}. The background oxygen concentration is peaked at the surface of the wafer due to surface contamination and post-growth formation of native oxides. In addition, there is an oxygen peak near the Si QW, presumably due to the switching of the mass flow controllers in the CVD reactor. In Fig.~\ref{fig:2}(d) we plot the concentration of oxygen atoms at the QW, $N_\text{o}$, for each wafer. For wafers 1 to 11, $N_\text{o}$ decreases from 8.7 $\times$ $10^{18}/$cm$^{3}$ to the SIMS detection threshold of 1 $\times$ $10^{17}/$cm$^{3}$. The decrease in oxygen concentration is correlated with the increase in mobility observed in Fig.\ 2(b--c). Wafer No.\ 26, which marks the beginning of a second cassette of wafers, shows an abrupt increase in $N_\text{o}$, which is also correlated with a drop in the mobility. The combination of mobility and SIMS data suggest that oxygen contamination is a mobility limiting factor in these undoped Si/SiGe heterostructures. A similar correlation has been observed in undoped Si/SiGe heterostructures grown by molecular beam epitaxy \cite{CJKRichardson}.

In addition to the correlation between $N_\text{o}$ and $\mu$, the data show that the heterostructure growth profile impacts the mobility of samples later in the growth series. As $h$ is increased from 40 to 50 nm for Wafers No.\ 15 and 16, we observe a corresponding increase in $\mu_\text{H}$ and $\mu_\text{L}$. For wafers 16--19, $h$ is constant and both $\mu_\text{H}$ and $\mu_\text{L}$ show very little variation. At Wafer No.\ 20, $h$ undergoes a large decrease from 50 to 20 nm, which is correlated with a large drop in mobility. For wafers 20--25, $h$ increases from 20 to 50 nm and we see that the mobilities also recover to the values obtained from Wafer No.\ 19. It is also notable that the correlation between $h$ and $\mu_\text{L}$ is stronger than that between $h$ and $\mu_\text{H}$, based on the relative sizes of the dips at Wafer No.\ 20. In contrast, for lower wafer numbers, the correlations between $h$ and mobility are weaker, suggesting that $N_\text{o}$ is the dominant mobility-limiting mechanism in these wafers. To obtain a better quantitative understanding of these correlations, we perform detailed measurements on Wafer No.\ 5 and 16 at $T = 0.35$ K in order to contrast the properties of a low and high mobility wafer.

\section{High mobility sample}

\begin{figure}
\centering
\includegraphics[width=\columnwidth]{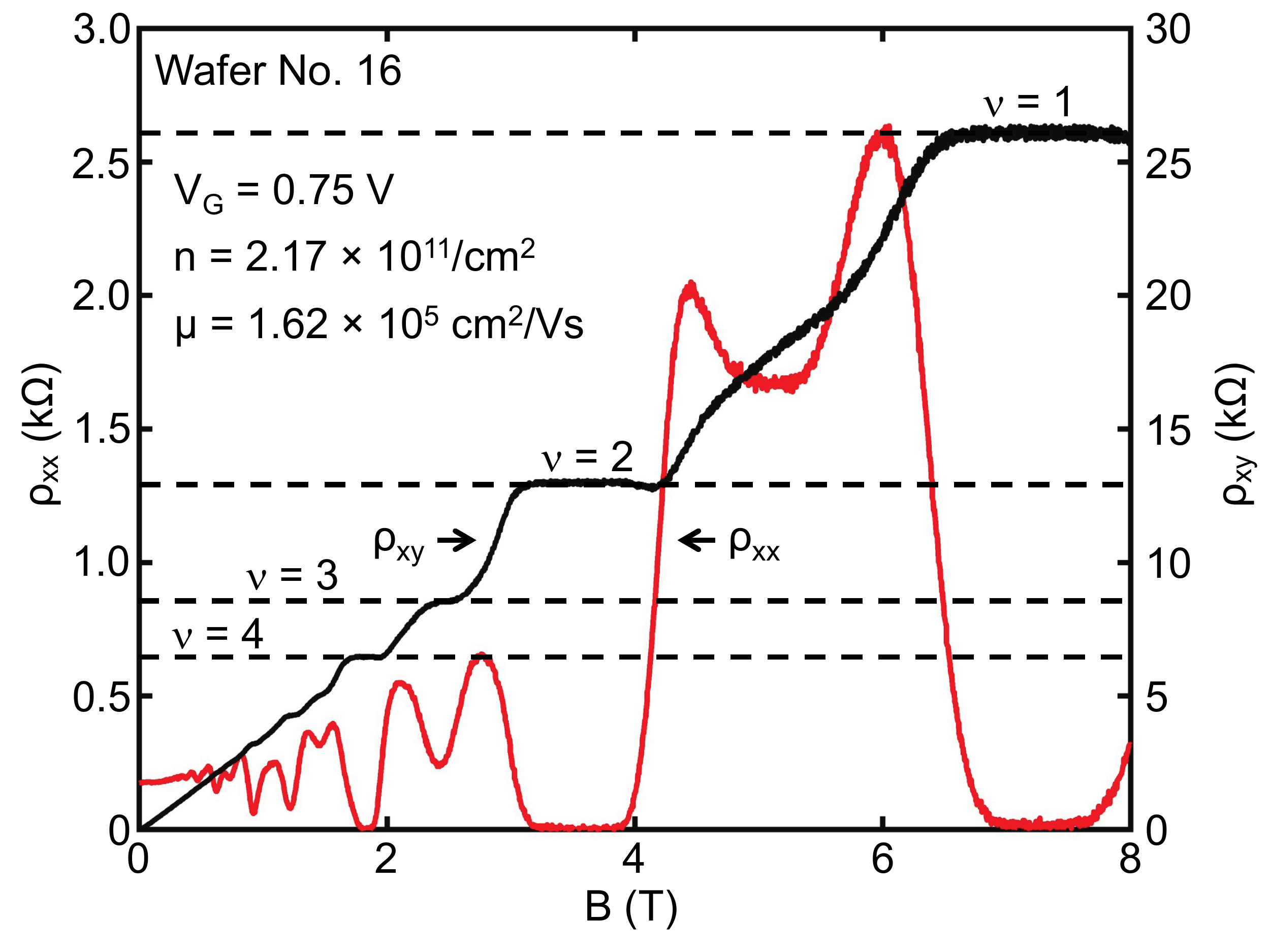}
\caption{Wafer No.\ 16. $\rho_\text{xx}$ (red) and $\rho_\text{xy}$ (black) as a function of $B$, with $V_{\text G} = 0.75$ V and $T = 0.35$ K. $n$ and $\mu$ are extracted from the resistivity values at $B$ = 0.1 T. Shubnikov-de Haas (SdH) oscillations are visible at intermediate fields 0.3 T $< B <$ 2 T. We observe clear quantum Hall plateaus in $\rho_\text{xy}$ at integer filling factors $\nu$ for $B >$ 1.5 T.}
\label{fig:3}
\end{figure}

Based on its high 4.2 K mobility, a Hall bar from Wafer No.\ 16 was cooled down in a $^3$He cryostat for further study. The oxygen content at the QW is $N_\text{o} = 0.5 \times 10^{18}/$cm$^3$ \cite{Supplement}. Figure~\ref{fig:3} shows characteristic plots of $\rho_\text{xx}$ and $\rho_\text{xy}$ as functions of $B$ up to 8 T, with $n = 2.17 \times 10^{11}/$cm$^{2}$. From the low field magnetotransport data we extract $\mu = 1.62 \times 10^5$ cm$^2$/Vs. We observe quantum Hall plateaus in $\rho_\text{xy}$ at consecutive integer filling factors $\nu$ for $B >$ 1.5 T, which indicates that both spin and valley degeneracies are lifted. In addition, $\rho_\text{xx}$ displays clear zeros, ruling out parallel conduction paths. For $\nu > 6$, plateaus in $\rho_\text{xy}$ are no longer visible, although oscillations in $\rho_\text{xx}$ are visible up to $\nu = 24$.

\begin{figure}
\centering
\includegraphics[width=\columnwidth]{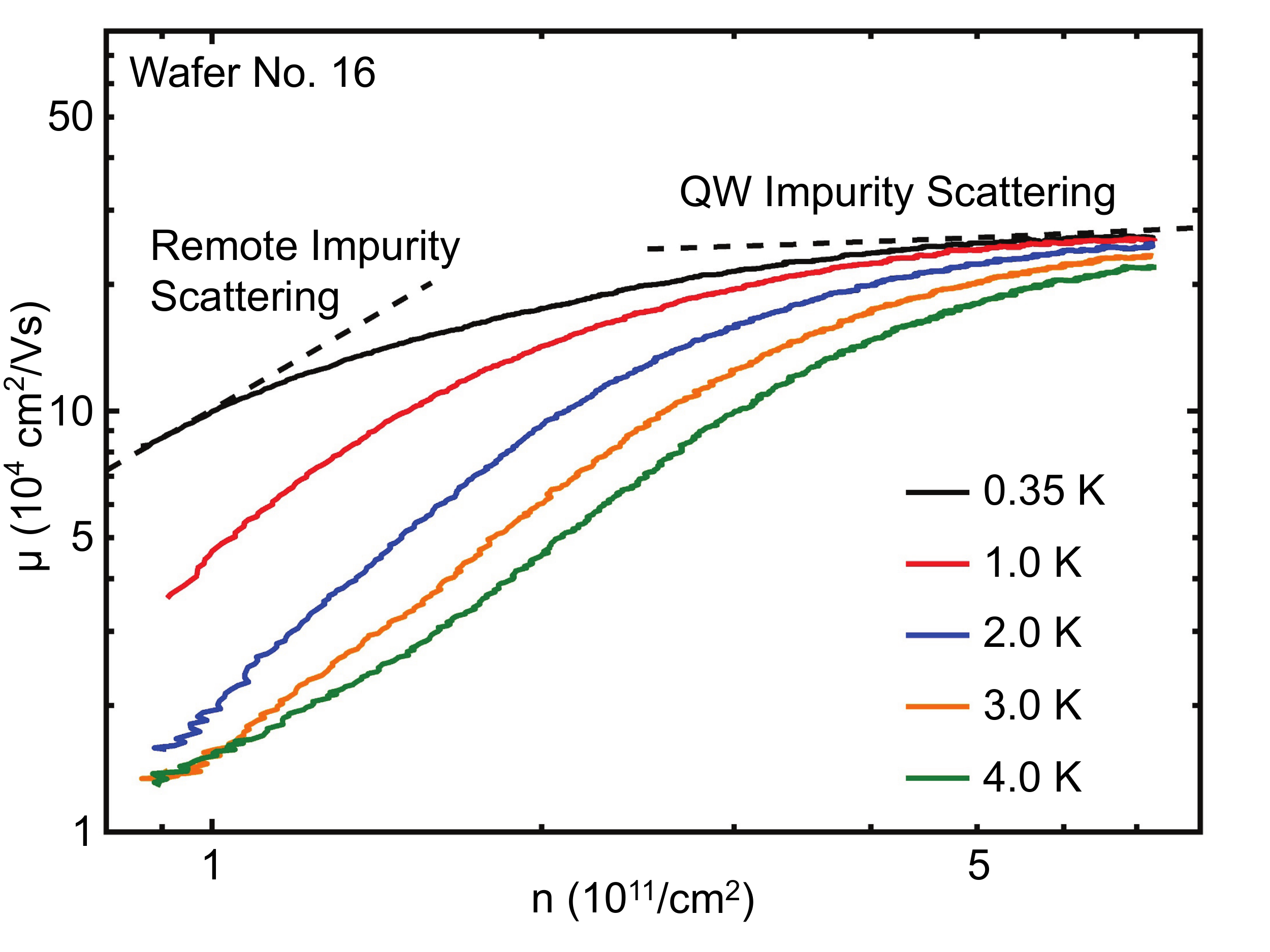}
\caption{Wafer No.\ 16. $\mu$ as a function of $n$ for five different temperatures. The dashed lines show the predicted slopes for remote ionized impurity scattering, which obeys a power law dependence $\mu \propto n^{1.5}$, and scattering due to impurities in the QW, which scales as $\mu \propto n^{0.1}$ on average \cite{Monroe_Mobility}. The charged impurity densities used to produce the two dashed lines are 3.7 $\times 10^{12}$/cm$^{2}$ at a distance of 50 nm from the QW center (where the Al$_2$O$_3$/Si interface is) for remote impurity scattering, and 3.4 $\times 10^{9}$/cm$^{2}$ in the QW center.}
\label{fig:4}
\end{figure}

For a single dominant scattering mechanism, the electron mobility is expected to scale as a power law of density \cite{Monroe_Mobility}, $\mu \propto n^{\alpha}$, with a scattering-mechanism-dependent exponent $\alpha$. Figure~\ref{fig:4} shows $\mu$ as a function of $n$ at five temperatures ranging from 0.35 to 4 K. At $T$ = 0.35 K, $\mu (n)$ is not well described by a single exponent, an often observed feature in 2DEG systems \cite{Jiang_GaA_Mobility}. Our data differ from previous work \cite{Sturm_APL2012}, where an exponent of $\alpha = 1.7$ is observed in the density range of $n = 0.6 \times 10^{11}$ /cm$^2$ to $n = 4.5 \times 10^{11}$ /cm$^2$. For $n$ $<$ $1 \times 10^{11}$/cm$^{2}$, the data roughly follow the $\mu \propto n^{1.5}$ scaling, which is consistent with scattering due to remote charged impurities \cite{Monroe_Mobility}. At higher $n$, $\mu$ increases at a much slower rate and displays signs of saturation when $n > 5 \times 10^{11}/$cm$^2$. The high density saturation likely arises from impurity charges located very near or inside the QW, which lead to values of $\mu$ that are only weakly dependent on $n$ \cite{Monroe_Mobility}. It is notable that the mobility curves are temperature dependent at low densities, but all saturate to nearly the same high density value of 250,000 cm$^{2}$/Vs. Another feature of the higher temperature data is that the density dependence of $\mu (n)$ becomes stronger, though the curvature persists up to 4 K. At $T = 4$ K, the data approximately follow a $\mu \propto n^{1.5}$ trend for  $n$ $<$ 3 $\times 10^{11}$/cm$^{2}$.

\begin{figure}
\centering
\includegraphics[width=\columnwidth]{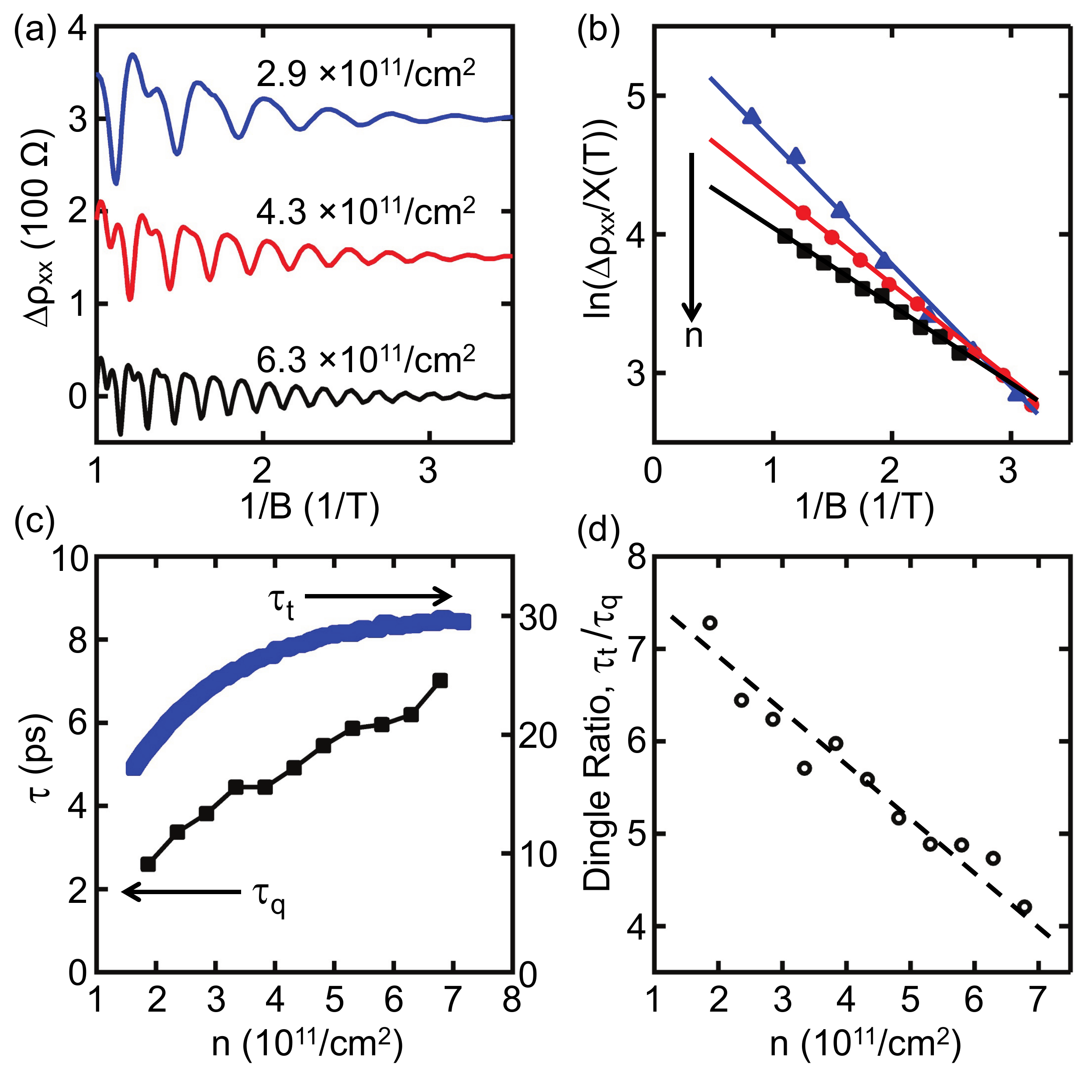}
\caption{Wafer No.\ 16. (a) $\Delta \rho_\text{xx} = \rho_\text{xx} - \rho_\text{b}$ as a function of $1 / B$ for three electron densities at $T = 0.4$ K, showing clear SdH oscillations. Traces have been offset by 150 $\Omega$ for clarity. (b) Dingle plots for the data in (a). Values of $\Delta \rho_\text{xx}$ used in this plot are the average of the maximum and minimum of each period of the SdH oscillations shown in (a). The SdH oscillations decay more slowly at higher densities. (c) Quantum lifetime $\tau_{\text q}$ and transport lifetime $\tau_{\text t}$ as functions of $n$ at $T = 0.4$ K. (d) Dingle ratio $\tau_{\text t} / \tau_{\text q}$ as a function of $n$ at $T = 0.4$ K. The dashed line is a guide to the eye showing a linear trend in the data.}
\label{fig:5}
\end{figure}

To further probe the scattering mechanisms that limit the mobility of Wafer No.\ 16, we measure low-field SdH oscillations in the longitudinal resistivity, $\rho_\text{xx}$. To facilitate the extraction of quantum lifetimes, we subtract the slowly varying background from $\rho_\text{xx}$ as outlined by Coleridge \textit{et al.} \cite{Coleridge_InterSubband} yielding $\Delta \rho_\text{xx} = \rho_\text{xx} - \rho_\text{b}$. Here $\rho_\text{b}$ is a slowly varying, polynomial background that has no detectable oscillatory component in the field range studied, indicating the lack of any appreciable parallel conduction path or inter-subband scattering in our sample \cite{Coleridge_InterSubband,Coleridge_SmallAngle}. $\Delta \rho_\text{xx}$ is plotted against $1/B$ for three densities in Fig.\ \ref{fig:5}(a). Clear periodic oscillations are observed, with a periodicity of 4 in $\nu$. This is consistent with the 2-fold spin degeneracy and 2-fold valley degeneracy at low fields. At higher fields $B > 0.7$ T, splitting of the peak in each period of the SdH oscillation becomes visible, which is a consequence of the increased Zeeman splitting which breaks the spin degeneracy of each Landau level. This splitting is examined in detail in Section VIII. We extract the amplitude of the oscillations in $\Delta \rho_\text{xx}$ at each period in $1/B$ using the method of linear interpolation outlined by Padmanabhan {\it et al.} \cite{Shayegan_EffectiveMassSuppression}. The decay in the resulting $\Delta \rho_\text{xx}$ amplitudes is fit according to \cite{Coleridge_SmallAngle}:
\begin{equation}
\Delta \rho_\text{xx} = 4 \rho_0 X(T) \exp (-\pi / \omega_\text{c} \tau_\text{q})
\end{equation}
where $\tau_\text{q}$ is the quantum lifetime of the electrons, $\rho_\text{0}$ is the zero-field resistivity, $X(T) = (2 \pi^2 k_\text{B} T / \hbar \omega_\text{c}) / \sinh (2 \pi^2 k_\text{B} T / \hbar \omega_\text{c})$ is the temperature-damping factor,
$\omega_\text{c} = eB / m^*$ is the cyclotron frequency, and $k_{B}$ is the Boltzmann's constant. We use a constant effective mass $m^* = 0.2 m_\text{e}$, where $m_\text{e}$ is the free electron mass, for all fits \cite{Schaffler_SiGe_Review}. Figure~\ref{fig:5}(b) shows the results of such fits, known as Dingle plots. The slopes of the Dingle plots \cite{Coleridge_SmallAngle} are inversely proportional to the quantum lifetime $\tau_\text{q}$.  We observe an increasing slope at lower electron densities, which suggests shorter quantum lifetimes at lower densities. We note that the Dingle plots show linear trends, suggesting the absence of any appreciable density inhomogeneities in our samples which would otherwise introduce uncertainties into estimates of $\tau_\text{q}$ \cite{Coleridge_InterSubband,Coleridge_SmallAngle}.

In Fig.~\ref{fig:5}(c), we compare the transport lifetime, $\tau_\text{t}$, and the quantum lifetime, $\tau_\text{q}$, across the electron density range $n$ = 1.8--6.8 $\times$ $10^{11}/$cm$^{2}$. Values of $\tau_\text{t}$ are obtained \cite{ashcroft1976solid} from the mobility data in Fig.~\ref{fig:4} via $\tau_\text{t} = \mu m^* / e$ and values of $\tau_\text{q}$ are obtained from analysis of the low field SdH oscillations. Both lifetimes show similar dependencies on $n$ and Dingle ratios, defined as $\tau_\text{t} / \tau_\text{q}$, range from 4 to 7 as shown in Fig.~\ref{fig:5}(d). In comparison with previous work on GaAs/AlGaAs \cite{Coleridge_SmallAngle}, GaN/AlGaN \cite{APL_GaN_2004} and modulation-doped SiGe \cite{DopedSiGe_APL_1995} where the Dingle ratios typically range from 10 to 20, the Dingle ratios measured for this undoped sample are sizably smaller. The relatively small Dingle ratio indicates that large angle scattering plays a more dominant role in this sample than in these traditional systems, a situation which would arise when the distribution of impurities is more concentrated towards the location of the 2DEG \cite{Monroe_Mobility}. Such a distribution is contrary to what is expected in an undoped system where charged impurities are thought to reside mostly in the Al$_2$O$_3$/Si interface \cite{Sturm_APL2012}, $\sim$50 nm away from the 2DEG in this sample. Our interpretation of the possible cause for such distribution is the peak in oxygen impurities at the 2DEG location \cite{Supplement}. Ionization of a small fraction of these oxygen atoms would lead to a sizable amount of impurity charges inside the QW, which contribute to large angle scattering with a Dingle ratio near unity. The decreasing trend of the Dingle ratio at higher densities also differs from theoretical calculations based on a single dopant sheet \cite{DasSharma_PRB1985} and previous work on a AlGaN/GaN system \cite{APL_GaN_2004}. This deviation is interpreted to be due to the mitigated contribution of remote impurity scattering to the overall momentum scattering rate at higher densities, since the scattering rate $\tau_\text{t}^{-1} \propto n^{-1.5}$ for remote impurities but $\tau_\text{t}^{-1} \propto n^{-0.1}$ for impurities inside the QW. At higher densities, scattering from impurities inside the QW becomes more dominant than remote impurities, which reduces the overall Dingle ratio.
\begin{figure}
\centering
\includegraphics[width=\columnwidth]{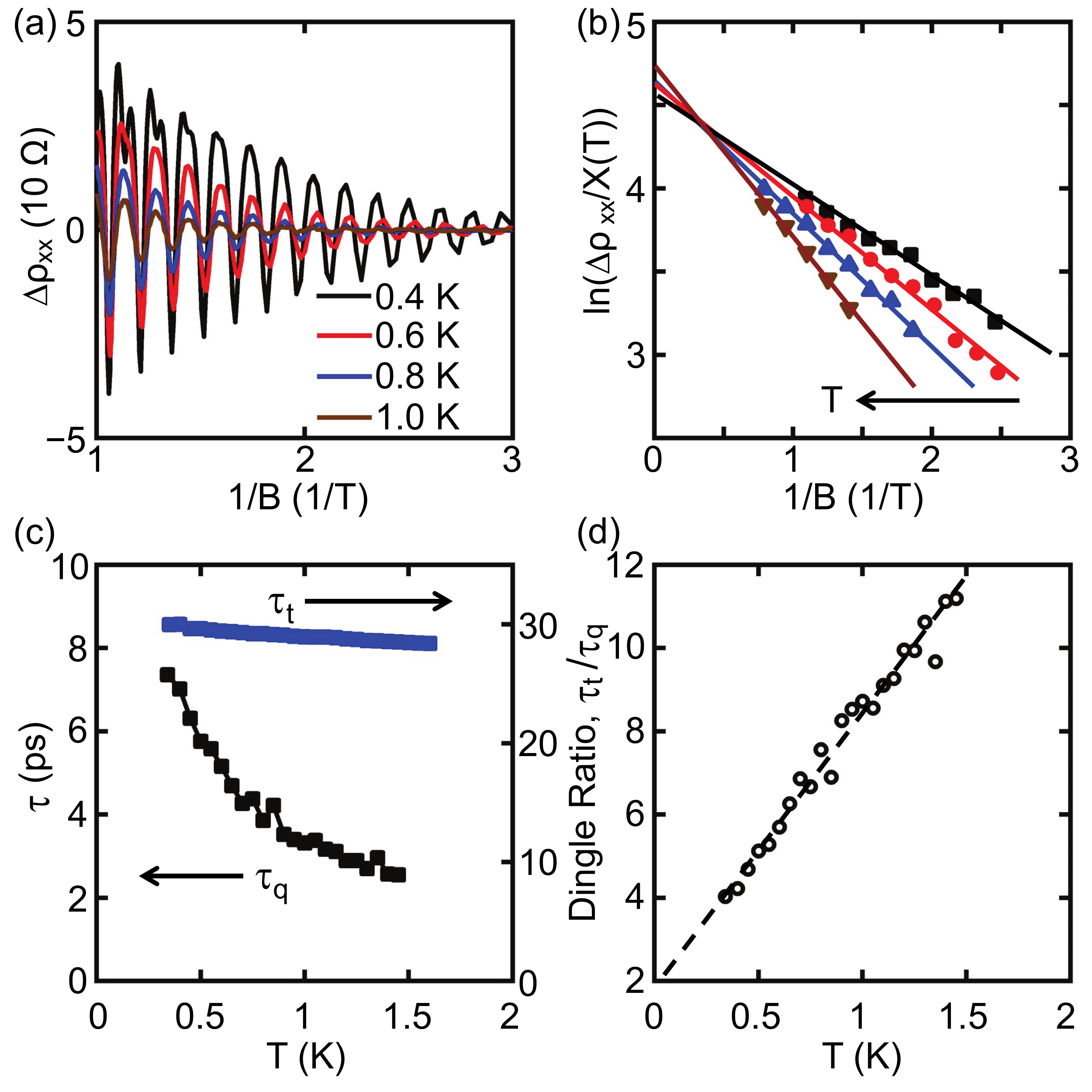}
\caption{Wafer No.\ 16. (a) $\Delta \rho_\text{xx}$ as a function of $1 / B$ for $n$ = 6.7 $\times 10^{11}$/cm$^{2}$ at four different temperatures. (b) Dingle plots for the data shown in (a). The SdH oscillations decay more rapidly at higher temperatures. (c) $\tau_{\text q}$ and $\tau_{\text t}$ as a function of $T$ at $n$ = 6.7 $\times 10^{11}$/cm$^{2}$. (d) Dingle ratio as a function of $T$. The dashed line is a guide to the eye.}
\label{fig:6}
\end{figure}

The quantum lifetime $\tau_\text{q}$ for this high mobility sample is also measured as a function of $T$ at $n$ = 6.7 $\times 10^{11}/$cm$^{2}$. Figure~\ref{fig:6}(a--b) displays low field SdH oscillations and the associated Dingle plots. The extracted values of $\tau_\text{q}$ are plotted alongside $\tau_\text{t}$, obtained by measuring the temperature-dependent mobility at this density [Fig.~\ref{fig:6}(c)]. The transport lifetime is relatively insensitive to temperature, while the quantum lifetime varies by nearly a factor of 4 from $T = 0.4$ K to $1.5$ K. The resulting Dingle ratio, plotted in Fig.~\ref{fig:6}(d), increases almost linearly with temperature from 4 to 11. These data are in contrast with a single-particle description of electron scattering in 2DEGs \cite{DasSharma_SiGe_PRB2005,2DEG_ClassicReview,DasSharma_PRB1985}, where the temperature dependence of both lifetimes is expected to be weak in the $T \ll T_\text{F}$ regime, where $T_\text{F}$ is the Fermi temperature (approximately 47 K at this density). Arapov {\it et al.} \cite{InGaA_GaA_TempDependent_Dingle} have recently measured an InGaA/GaAs double QW structure and report a similar, strong temperature dependence of $\tau_\text{q}$ at $T \ll T_\text{F}$, which the authors attribute to electron-electron interactions. A similar mechanism may explain the trends observed in this work.

\section{Low Mobility sample}

\begin{figure}
\centering
\includegraphics[width=\columnwidth]{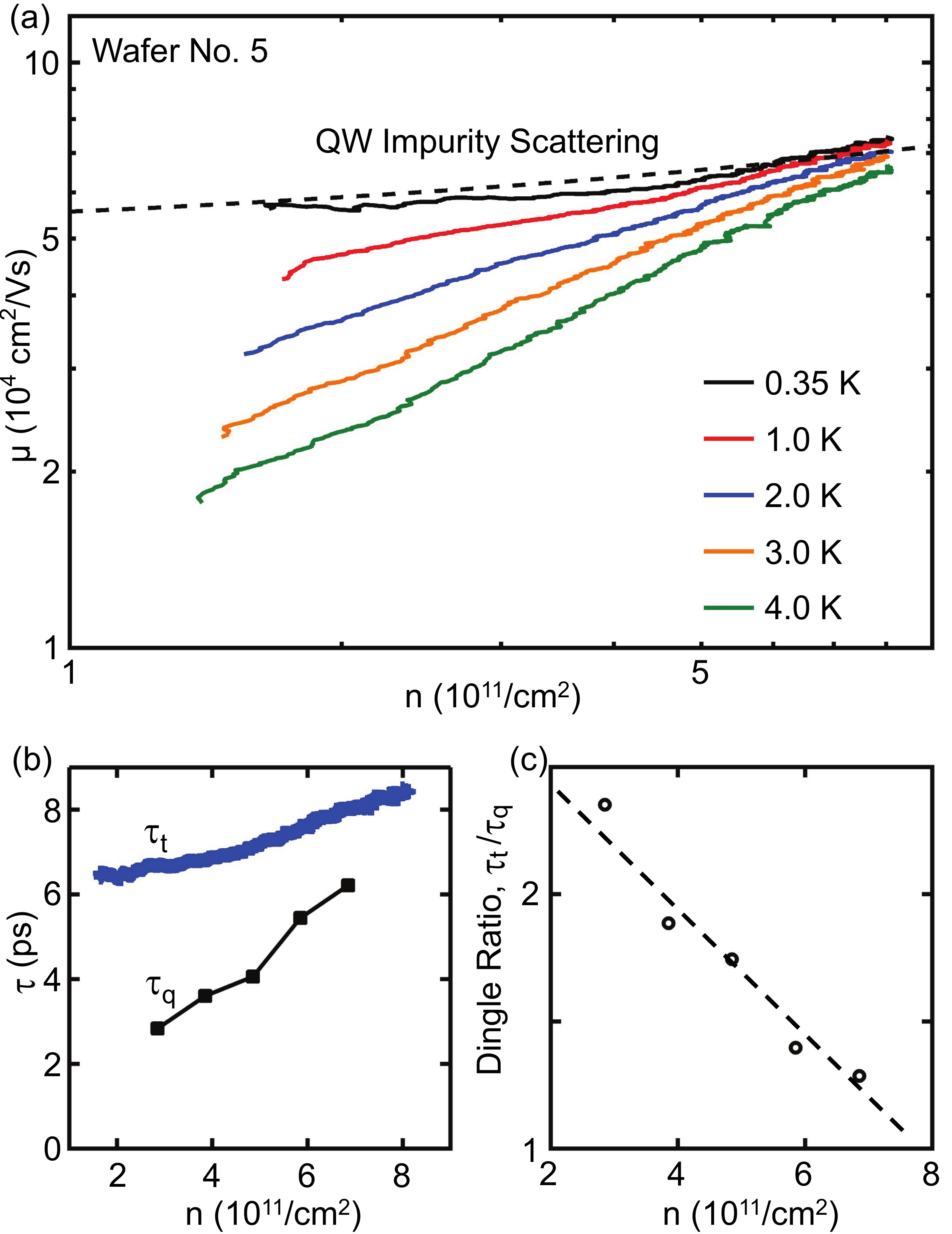}
\caption{Wafer No.\ 5 (low mobility sample). (a) $\mu(n)$ at five different temperatures. The dashed line shows scattering due to impurities in the QW ($\mu \propto n^{0.1}$ on average). The charged impurity density used to produce the dashed line is 1.3 $\times 10^{10}/$cm$^2$ at the QW center. (b) $\tau_\text{q}$ and $\tau_\text{t}$ plotted as a function of $n$, extracted from low field SdH oscillations. $T = 0.4$ K in this plot. (c) Dingle ratios $\tau_\text{t} / \tau_\text{q}$ obtained from the data in (b).}
\label{fig:7}
\end{figure}

We next examine data from Wafer No.\ 5, which has a much lower maximum mobility of $\mu$ = 7.5 $\times$ $10^4$ cm$^2/$Vs at 4.2 K. Wafer No.\ 5 has a 2 nm thick Si cap, a $h$ = 40 nm thick SiGe spacer layer, and a 5 nm wide Si QW. SIMS analysis shows a similar distribution of oxygen inside the SiGe spacer as the high mobility sample \cite{Supplement}. However, the oxygen content in the QW is peaked at $N_\text{o}$ = 2.5 $\times$ $10^{18}/$cm$^{3}$, which is five times higher than the high mobility sample.

Figure~\ref{fig:7} shows the results of magnetotransport measurements on this low mobility sample. Reliable measurements of $\mu$ could only be performed on this sample for $n$ $>$ 1.7 $\times 10^{11}$/cm$^{2}$. Below this density the sample is in an apparent insulating state. $\mu(n)$ is plotted in Fig.~\ref{fig:7}(a) and increases with $n$, although with a weaker dependence than the high mobility sample. $\mu (n)$ is also temperature dependent, and more strongly scales with $n$ at higher temperatures. At $T = 0.35 $ K, $\mu (n)$ is nearly density-independent. As $T$ increases, $\mu (n)$ becomes more density-dependent and eventually reaches an approximate scaling of $\mu \propto n^{0.7}$ at $T$ = 4 K. The smaller power-law exponent for this sample suggests that remote impurity scattering plays a less significant role compared to the high mobility sample. Instead, electron scattering is likely dominated by impurity charges situated inside the QW, which is consistent with the higher oxygen content observed in the SIMS data \cite{Supplement}.

Figure \ref{fig:7}(b) shows $\tau_\text{q}$ and $\tau_\text{t}$ for the low mobility sample at five different densities and with $T = 0.4 $ K. Both lifetimes are shorter at lower electron densities, similar to the high mobility sample. Interestingly, despite the factor of $\sim$3 difference in $\tau_\text{t}$ between the two samples, the values of $\tau_\text{q}$ are very similar. This observation agrees well with recent theoretical results by Das Sarma {\it et al.} \cite{PhysRevB.90.035425}, who considered a two-impurity model and showed that increasing (decreasing) $\tau_\text{t}$ does not necessarily lead to increasing (decreasing) $\tau_\text{q}$ when there is more than one scattering mechanism. We also plot the density-dependent Dingle ratio $\tau_\text{t}/\tau_\text{q}$ of this sample in Fig.\ \ref{fig:7}(c). $\tau_\text{t}/\tau_\text{q}$ ranges from 1.3 at high density to 2.3 at low density, significantly smaller than the high mobility sample. We were only able to measure $\tau_\text{q}$ down to a density of 2.85 $\times 10^{11}/$cm$^2$ in this sample due to its relatively lower quality which makes clear Dingle plots difficult to obtain at lower densities. The smaller Dingle ratio observed in this sample implies that the underlying scattering events are even larger in angle compared to the high mobility sample, consistent with scattering from QW impurities.

\section{Estimate of Defect Densities}
Monroe {\it et al.} have carefully analyzed seven scattering mechanisms  that are potentially relevant to the Si/SiGe materials system  \cite{Monroe_Mobility}. Among these mechanisms, alloy scattering,  scattering due to strain modulation, scattering due to vicinal surfaces  and scattering from threading dislocations are estimated to limit  mobilities to above $10^7$ cm$^{2}$/Vs, two orders of magnitude higher  than the mobilities measured in our samples. Interface roughness has  been reported to be an important factor in a $\sim$ 500 nm deep Si/SiGe QW structure \cite{Tsui_APL2012_HighSiGeMobility},  but is expected to lead to a mobility that decreases with increasing  density (a trend that is not observed in our data). We therefore limit  our analysis to the two remaining scattering mechanisms: remote impurity scattering and scattering from background charges.

In this section we first review the theory of Monroe {\it et al.} \cite{Monroe_Mobility}. We  then compare the measured mobility and Dingle ratio with predictions  from this theory, allowing us to estimate charged defect densities in  the high and low mobility samples. We limit our analysis to the $T =  0.35$ K data, where $T \ll T_\text{F}$ is satisfied throughout the  density range studied in Fig.~\ref{fig:4} and Fig.~\ref{fig:7}(a), and  thermal effects are negligible.

Remote impurity scattering is often identified as the dominant mobility-limiting mechanism in doped Si/SiGe heterostructures, where the  dopant atoms contribute to the formation of a sheet of disordered  charges located a setback distance $z_0$ from the 2DEG \cite{DopedSiGe_APL_1995,Sugii_DopedSiGe_1998,Sturm_APL2012}. These disordered charges result in potential fluctuations in the QW, contributing to electronic scattering and reduced mobility.  The functional form that relates mobility to the electron density and 2D  density of disordered charges $n_1$ was derived by Monroe {\it et al.}:
\begin{equation}
\mu \approx 16 \pi^{1/2} g_\text{v}^{1/2} g_\text{s}^{1/2} e n^{3/2}  z_0^{3} / \hbar n_1
\end{equation}
where $g_\text{v} = 2$ ($g_\text{s} = 2$) accounts for the valley (spin)  degeneracy. Remote impurity scattering results in a relatively strong  density dependence $\mu(n) \propto n^{1.5}$, which is attributed to the fact that at higher $n$, the Fermi wavevector $k_\text{F}$ increases and electrons are scattered through smaller angles by the potential fluctuations of the same remote impurities. Monroe \textit{et al.} also make a prediction for the density dependence of the Dingle ratio for remote impurity scattering:
\begin{equation}
\tau_\text{t} / \tau_\text{q} \approx (16 \pi / g_\text{v} g_\text{s})  z_0^{2} n.
\end{equation}
Thus, theory predicts $\tau_\text{t} / \tau_\text{q} \propto n$ and  $\tau_\text{t} / \tau_\text{q} \propto z_0^{2}$ respectively. Monroe {\it et al.} obtained the closed-form expressions of Eqns. (2) and (3) based on the assumption that $k_\text{F} z_0 \ge 1$. Since $k_\text{F} \approx 5.6 \times 10^5$ /cm at $n = 1 \times 10^{11}$ /cm$^2$, this assumption translates to a setback distance $z_0 \ge 18$ nm.

For the case of scattering due to more arbitrarily distributed background charges where $k_\text{F} z_0 \ge 1$ does not necessarily hold, Monroe {\it et  al.} derived a general expression for $\tau_\text{t}$ (related to $\mu$  via $\tau_\text{t} = \mu m^* / e$):
\begin{equation}
\tau_\text{t}^{-1} = \frac{m^*}{\pi \hbar^3} \frac{1}{k_\text{F}}  \int_{0}^{2 k_\text{F}} dq \frac{S(q)}{\sqrt{1 - q^2 / 4k_\text{F}^2}}  \frac{q^2}{2k_\text{F}^2}
\end{equation}
where $k_\text{F} = \sqrt{4 \pi n / g_\text{s} g_\text{v}}$. The integration variable $q$ physically  represents the magnitude of the change in wavevector for a given scattering event. The expression for $\tau_\text{q}$ is similar but  without the angle-weighing factor $\frac{q^2}{2k_\text{F}^2}$:
\begin{equation}
\tau_\text{q}^{-1} = \frac{m^*}{\pi \hbar^3} \frac{1}{k_\text{F}}  \int_{0}^{2 k_\text{F}} dq \frac{S(q)}{\sqrt{1 - q^2 / 4k_\text{F}^2}}. \end{equation}
$S(q)$ is the power spectral  density of impurity charges:
\begin{equation}
S(q) = \frac{e^4}{[2 \epsilon_r \epsilon_0 (q + q_\text{s})]^2}  \int_{-\infty}^{\infty} dz N_\text{i}(z) \exp(-2q |z|)
\end{equation}
where $q_\text{s} = e^2 g_\text{2D} / (2 \epsilon_r \epsilon_0)$ is the  Thomas-Fermi screening wavevector, $\epsilon_r = 12$ is the relative  permittivity of Si and $g_\text{2D} = (g_\text{v} g_\text{s} / 2 \pi)  (m^* / \hbar^2)$ is the 2D density of states. $N_\text{i} (z)$ is the 3D  density of impurity charges, where $z$ is measured relative to the  center of the 2DEG wavefunction ($z = 0$). Based on this expression, the special case of a  uniform background charge $N_\text{i} (z) = N_\text{b}$ gives $\mu =  (g_\text{v}^{3/2} g_\text{s}^{3/2} / 4 \pi^{1/2}) (en^{1/2} / \hbar  N_\text{b})$. The Dingle ratio for a uniform background charge density was not explicitly derived by Monroe {\it et al.}, but is expected to be large.

Considering the SIMS analysis, which shows the presence of oxygen in the  QW, we also analyze scattering from a 2D sheet of charged impurities with density $n_2$ located at $z = 0$, i.e. $N_\text{i} (z) = n_2 \delta  (z)$. This case was not explicitly analyzed by Monroe \textit{et al.} We therefore numerically  integrate Eqns.\ (4--5) to obtain $\mu$ for given values of $n_2$ and  $n$. The resulting density dependence $\mu(n)$ is very weak, with $\mu  \propto n^{0.1}$ on average. We also find that the Dingle ratio  $\tau_\text{t} / \tau_\text{q} \approx 1$ for $n$ $>$ 0.9 $\times 10^{11}/$cm$^2$.

We now compare the experimental data with these predictions, starting  with the high mobility sample. Figure 4 shows $\mu(n)$ for Wafer No.\  16. At low densities, $\mu(n)$ roughly follows the power law expected  for remote impurity scattering, while for higher densities $\mu$ is a  weak function of $n$. Superimposed on the data are dashed lines showing  the expected scaling for remote impurity scattering and scattering from  impurities in the QW. To compare with theory for remote  impurity scattering, we set $z_0$ = 50 nm which is the SiGe spacer thickness $h$ of this sample such that the remote impurities are Al$_2$O$_3$/Si interface charges, as reported by Li {\it et al.} \cite{Sturm_APL2012}. We then adjust $n_1 = 3.7 \times 10^{12}$ /cm$^2$ to bring theory into agreement with the data. Similarly, for scattering from impurities in the QW we find reasonable agreement with the data when $n_1$ = 3.4 $\times 10^{9}/$cm$^2$. Dingle ratio data for Wafer  No.\ 16 are plotted in Fig.~\ref{fig:5}(d) and show a linear decrease with $n$ over the entire density range. This is broadly consistent with a crossover from remote impurity scattering limited transport to local defect scattering-limited transport as $n$ increases.

In comparison, $\mu(n)$ is shown for the low mobility sample (Wafer No.\  5) in Fig.\ 7(a). At $T$ = 0.35 K, the mobility is weakly dependent on  density over the entire density range, consistent with scattering from  impurities in the QW. The dashed line shows  the prediction for scattering from impurities in the QW taking  $n_2$ = 1.3 $\times 10^{10}/$cm$^2$. We note that this defect density is 4 times higher than the high mobility sample, reminiscent of the  factor of 5 difference between the oxygen contents in the QWs of  the low and high mobility samples \cite{Supplement}. It is also clear  that the Dingle ratio is much less sensitive to density, with  $\tau_t/\tau_q$ $\sim$ 1 -- 2 over the entire density range. The small  Dingle ratio is consistent with scattering from impurities in the QW.

\section{Metal-to-Insulator Transition}

For spin-based quantum information processing, quantum dots are typically operated in the few-electron regime \cite{HRL_Nature2012}. It is therefore important to characterize the strength of the disorder potential in the low electron density regime. One important gauge for the degree of disorder is the critical electron density, $n_\text{c}$, for the metal-to-insulator transition (MIT) in 2DEGs. Recent experiment and theory have established the MIT as a percolation phenomenon, where a fraction of electrons become localized by the disorder potential \cite{PhysRevB.79.235307,DasSarma_MIT_Semiclassical}. As such, $n_\text{c}$ is an important gauge of the degree of disorder present in the QW system, and higher quality samples have been demonstrated to display lower values of $n_\text{c}$ \cite{Lai_MIT_SiGe_APL2004}. Achieving low $n_\text{c}$ is therefore important for few-electron quantum dots, since gate control of electrons may be compromised if the disorder potential is large. For silicon, a MIT was first observed in MOSFETs \cite{SiMOSFET_MIT_PRB1994,SiMOSFET_MIT_PRB1995,SiMOSFET_PRL1997,PhysRevB.79.235307,SiMOS_MIT_PRL_2008}, subsequently in modulation doped Si/SiGe structures \cite{Olshanetsky_DopedSiGe_MIT_PRB2003,Lai_MIT_SiGe_APL2004,Lai_SiGe_MIT_PRB2005} and more recently, an undoped Si/SiGe 2DEG structure \cite{Lai_MIT_Undoped_SiGe_APL2007}, and an ambipolar Si-vacuum FET \cite{2015arXiv150202956H}. In particular, values of $n_\text{c}$ vary greatly in Si/SiGe systems, ranging from 0.32 to 4.05 $\times 10^{11}/$cm$^{2}$ \cite{Olshanetsky_DopedSiGe_MIT_PRB2003,Lai_MIT_SiGe_APL2004,Lai_SiGe_MIT_PRB2005,Lai_MIT_Undoped_SiGe_APL2007}.

\begin{figure}[t]
\centering
\includegraphics[width=\columnwidth]{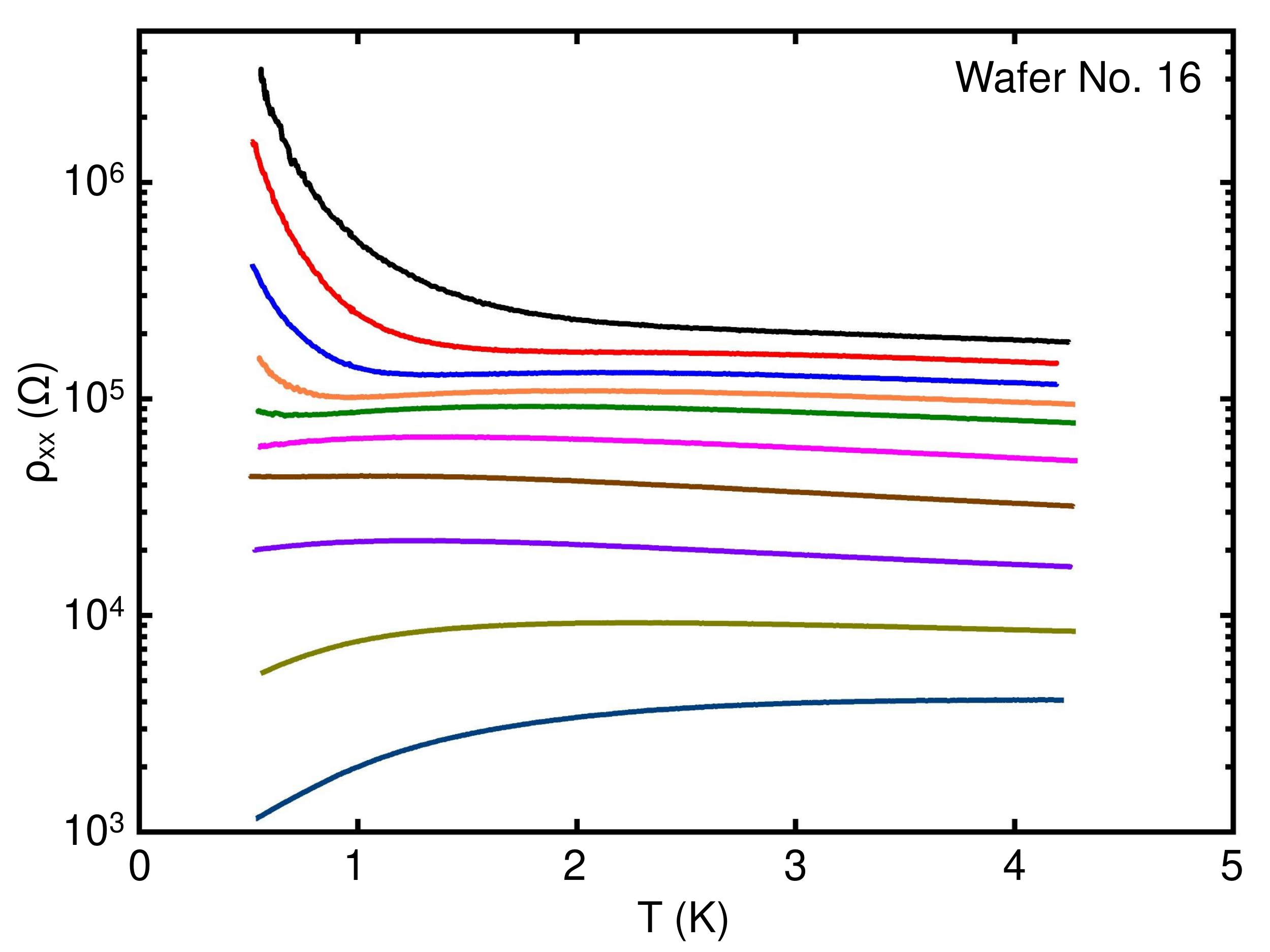}
\caption{Wafer No.\ 16. $\rho_\text{xx}$ as a function of $T$ at $B = 0$ T for $n =$ 0.34, 0.36, 0.38, 0.40, 0.42, 0.46, 0.51, 0.59, 0.73 and 0.96 $\times 10^{11}/$cm$^2$ (from top to bottom). A metal-insulator transition occurs at a critical density of $n_{\text c}$ = 0.46 $\times 10^{11}/$cm$^{2}$.}
\label{fig:8}
\end{figure}

The experimental signature for the MIT in 2DEG systems is a sign reversal in $d \rho / dT$, where $\rho$ is the resistivity of the system \cite{SiMOSFET_MIT_PRB1994}. For $n > n_\text{c}$, $d \rho / dT >$ 0 and the 2DEG displays metallic behavior. For $n < n_\text{c}$, $d \rho / dT <$ 0 and the 2DEG behaves as an insulator. In Fig.\ ~\ref{fig:8}, we plot $\rho_\text{xx}$ as a function of temperature for the high mobility sample at ten different densities below $n = 1.0 \times 10^{11}/$cm$^{2}$. We observe the following features in this data set:

1. At the lowest two densities $n = 0.34 \times 10^{11}$ /cm$^{2}$ and $0.36 \times 10^{11}$ /cm$^{2}$, $d \rho_\text{xx} / dT < 0$ throughout the measured temperature range. In addition, $\rho_\text{xx}$ appears to diverge exponentially at $T < 1$ K, indicative of a true insulating phase \cite{DasSarma_MIT_Theory_PRB2003}.

2. At the next three higher densities  $n =$ 0.38, 0.40 and 0.42 $\times 10^{11}/$cm$^{2}$, $\rho_\text{xx}$ varies non-monotonically with temperature. While $d \rho_\text{xx} / dT <$ 0 at $T = 4.2$ K, $d \rho_\text{xx} / dT >$ 0 for a small, intermediate temperature range. We note that this behavior has also been observed by Lu {\it et al.} in another undoped Si/SiGe sample \cite{Lai_MIT_Undoped_SiGe_APL2007}, and is common in Si MOSFET systems \cite{PhysRevB.90.125410}.

3. $d \rho_\text{xx} / dT >$ 0 at 0.5 K for $n \ge n_\text{c}$, where $n_\text{c} = 0.46 \times 10^{11}$/cm$^{2}$.

4. At $n \ge n_\text{c}$, with the exception of $n = 0.51 \times 10^{11}$ /cm$^{2}$, $d \rho_\text{xx} / dT >$ 0 up to a crossover temperature $T_\text{c}$. For $T > T_\text{c}$, $d \rho_\text{xx} / dT <$ 0. Furthermore, $T_\text{c}$ increases with increasing $n$. Das Sarma {\it et al.} interpreted this behavior as a quantum-to-classical crossover \cite{DasSarma_Si_MIT_PRL1999,DasSarma_MIT_Theory_PRB2003,PhysRevB.69.195305,DasSharma_SiGe_PRB2005}. We also find that at $T < 1.2$ K and $n = 0.96 \times 10^{11}$/cm$^{2}$ (such that $T \ll T_\text{F}$), $\rho_\text{xx}$ is well approximated by a power law relation $(\rho_\text{xx} - \rho_0) / \rho_0 = 14.8(T/T_\text{F}) + 30.7(T/T_\text{F})^{3/2}$ where $\rho_0$ is the zero-temperature resistivity with a value of 400 $\Omega$ based on linear extrapolation. This is also in good agreement with theoretical predictions \cite{PhysRevB.69.195305}.

We note that the critical density is comparable to the lowest value of 0.32 $\times 10^{11}/$cm$^{2}$ that has been reported in doped Si/SiGe structures \cite{Lai_SiGe_MIT_PRB2005}, and a factor of 4 lower than the value of 1.9 $\times 10^{11}/$cm$^{2}$ observed in a previous work on undoped Si/SiGe structure \cite{Lai_MIT_Undoped_SiGe_APL2007}, indicating a very low level of disorder in our undoped sample. More broadly, the critical density observed in our system lies at the lower end of the critical density spectrum \cite{MIT_Critical_Density_Review}. A low value of $n_\text{c}$ = 7.7 $\times 10^9/$cm$^{2}$ was obtained in the GaAs system \cite{GaAs_Record_Low_MIT}.

\section{Valley Splitting}

\begin{figure}
	\centering
	\includegraphics[width=\columnwidth]{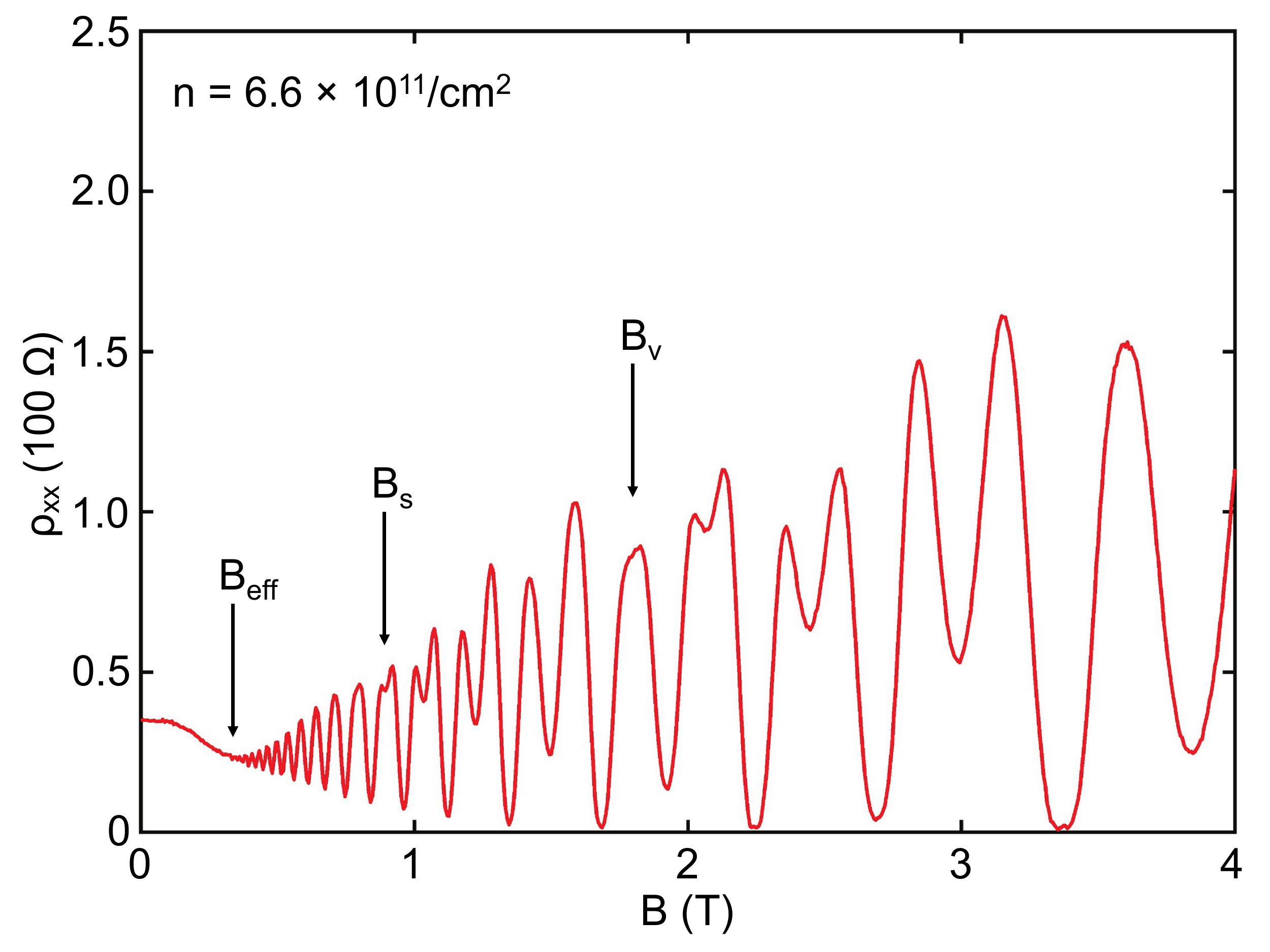}
	\caption{Wafer No.\ 16. $\rho_\text{xx}(B)$ at $T = 0.4$ K and $n = 6.6 \times 10^{11}$ /cm$^2$. The onset of SdH oscillations is observed at $B_\text{eff} = 0.38$ T. Spin degeneracy is lifted at $B_\text{s}$ = 0.88 T and the valley degeneracy is lifted at $B_\text{v}$ = 1.8 T.}
	\label{fig:9}
\end{figure}

 Another important figure of merit for the Si/SiGe quantum well system is the magnitude of the valley splitting. The conduction band of Si has six equivalent minima, or valleys. For Si/Si$_{0.7}$Ge$_{0.3}$ QWs, the in-plane strain increases the energies of the four in-plane valleys by $\sim$200 meV \cite{Monroe_Mobility,Schaffler_SiGe_Review,2DEG_ClassicReview}. The splitting of the two lowest lying valleys, $\Delta_\text{v}$, is dependent on the magnitude of the vertical electric field in the quantum well and the degree of disorder. Degeneracy of these valleys provides an additional route for spin decoherence. Measurements of the valley splitting in Si MOSFET systems have been extensively performed. Most values range from 0.7 -- 1.5 meV  \cite{Valley_SiMOSFET_1979,Nicholas198051,Wakabayashi1986359,Valley_JETP}, with one study reporting a value as large as 23 meV \cite{Giant_Valley}. In comparison, the valley splitting in Si/SiGe systems tends to be smaller, ranging from 0.05 -- 0.3 meV \cite{Weitz_SiGe_ValleySplitting,Koester_Valley_SiGe,Lai_SiGe_ValleySplitting_PRB2006,Eriksson_Valley_Nature2007}. In this section, we determine $\Delta_\text{v}$ through careful analysis of the SdH oscillations in Wafer No.\ 16.

In Fig.~\ref{fig:9} we plot $\rho_\text{xx}(B)$ with $n$ = $6.6 \times 10^{11}/$cm$^2$. SdH oscillations are observed above an effective field $B_\text{eff}$ = 0.38 T and have a periodicity of 4 in $\nu$. For $B$ $ > B_\text{s} =$ 0.88 T, we observe change in periodicity of the SdH oscillations, indicating that spin degeneracy has been lifted. The periodicity changes again beyond $B_\text{v}$ = 1.8 T, consistent with the lifting of both spin and valley degeneracies. We have verified that the spin degeneracy is lifted before valley degeneracy using the tilted field method \cite{Weitz_SiGe_ValleySplitting,Koester_Valley_SiGe,Lai_SiGe_ValleySplitting_PRB2006}.

The energy spectrum of 2D electrons in a perpendicular field is described by four characteristic energy scales. The first is the Zeeman splitting, $E_\text{z} = g \mu_\text{B} B$, where $\mu_\text{B}$ is the Bohr magneton and $g$ is the electronic $g$-factor. The second is $E_\text{l} = e \hbar B / m^* - E_\text{z}$, which is the Landau level spacing minus the Zeeman splitting. The third is the valley splitting, $\Delta_\text{v}$. Finally, clear SdH oscillations will only be observed when the Landau level spacing is greater than the Landau level broadening $\Gamma \approx \hbar / 2 \tau_q$. Spin splitting becomes visible when $E_\text{z} (B_\text{s}) \approx \Gamma$ . Based on the effective field at which the SdH oscillations become visible, we estimate $\Gamma \approx E_\text{l} (B_\text{eff})$. We then have the relation $E_\text{z} (B_\text{s}) \approx E_\text{l} (B_\text{eff})$, allowing us to extract $g = 3.02$. The g-factor is in reasonable agreement with the value of $g = 2.9 \pm 0.1$ at $n = 5.9 \times 10^{11}$ /cm$^2$ found in a previous study \cite{Koester_Valley_SiGe}. Based on this experimental value of $g$, we find $\Gamma$ $\sim$ 150 $\mu$eV. Finally, the valley degeneracy is lifted at the field for which $\Delta_\text{v} (B = 1.8 \text{ T})$ $\sim$ 150 $\mu$eV. This value for the valley splitting is substantial and comparable to the two-electron singlet-triplet splitting that is measured in GaAs quantum dots \cite{Petta_Science,Petta_RevMod}.

\section{Conclusions}
We have measured 26 wafers with different growth parameters to identify the dominant mobility limiting mechanisms in undoped Si/SiGe QW heterostructures. At 4.2 K we find correlations between mobility and oxygen content at the QW as well as the thickness of the top SiGe spacer. We have also measured the transport lifetime, $\tau_\text{t}$, and quantum lifetime, $\tau_\text{q}$, of two Si/SiGe QW heterostructures across a wide density range at $T$ $\sim$ 0.35 K. Based on the density dependencies of the two lifetimes, we conclude that the mobility of high quality samples with low oxygen content at the QW is mostly limited by remote impurity charges. Lower quality samples with high oxygen content at the QW are limited by the impurity charges inside or very close to the QW, consistent with the correlations observed at 4.2 K. To further assess the merits of the high mobility heterostructure as a platform for spin-based quantum dots, we have measured a low critical density $n_{\text c}$ = 0.46 $\times$ $10^{11}/$ cm$^{2}$ for the MIT and a valley splitting $\Delta_\text{v}$ $\sim$ 150 $\mu$eV.

While we cannot rule out effects due to other types of impurities, our SIMS results suggest that significant improvements in the electron mobility may be obtained by reducing the level of oxygen content in the Si/SiGe heterostructure, particularly in regions close to the QW. The SiGe spacer layer thickness can also be increased to reduce scattering from charged impurities at the surface of the wafer. This second approach has limitations for quantum dot devices, as it is desirable to have strong in-plane electrostatic confinement, which is harder to obtain in samples with deeper QWs. The magnetic fields at which the valley splitting was extracted corresponds to a cyclotron radius of $\sim$20 nm, which is a realistic size for the lithographic patterning of quantum dots on Si. Therefore efforts should also be directed towards reducing the size of Si quantum dots to emulate high levels of magnetic confinement, which yielded large values of valley splitting in this work. Overlapping gate architectures may prove helpful to achieve tight electronic confinement in the relatively high effective mass Si/SiGe quantum well system \cite{zajac2015}.

\begin{acknowledgements}
We thank S. Das Sarma and M. Shayegan for valuable discussions. Research sponsored by the United States Department of Defense with partial support from the NSF (DMR-1409556 and DMR-1420541). The views and conclusions contained in this document are those of the authors and should not be interpreted as representing the official policies, either expressly or implied, of the United States Department of Defense or the U.S. Government. Approved for public release, distribution unlimited.
\end{acknowledgements}

\end{document}